\newcommand{\be}{\begin{equation}}\newcommand{\ee}{\end{equation}}
\newcommand{\bea}{\begin{eqnarray}}\newcommand{\eea}{\end{eqnarray}}
\newcommand{\nn}{\nonumber}\newcommand{\p}[1]{(\ref{#1})}
 \newcommand{\lb}[1]{\label{#1}}
 \newcommand\q{\quad}
\newcommand\qq{\quad\quad}
\newcommand\s{\scriptscriptstyle}

\def\a{\alpha}

\def\b{\beta}

\def\g{\gamma}

\def\d{\delta}

\def\eps{\epsilon}

\def\ve{\varepsilon}

\def\j{\psi} \def\bj{{\bar\psi}}
  
\def\l{\lambda}

\def\o{\omega}

 \def\th{\theta}  \def\bt{\bar\theta}

\def\r{\rho}

\def\J{\Psi}

\def\L{\Lambda}

\def\pa{\partial}

\def\na{\nabla}

\newcommand\Tr{\mbox{Tr}\,}

\newcommand\tooS{(\theta^{(0,0)})^2}

\newcommand\ab{{\alpha\beta}}

\newcommand\A{{\s A}}

\newcommand\cD{{\cal D}}

\def\sfrac#1#2{{\textstyle\frac{#1}{#2}}}

\documentclass[12pt]{article}
\usepackage{amsfonts}
\begin{document}

\begin{center}
{\bf    GAUGE MODEL IN D=3, N=5 HARMONIC SUPERSPACE}\\
\vspace{0.5cm}
{\it
  B.M. Zupnik~\footnote{Talk at the International workshop "Supersymmetries
  and Quantum Symmetries", July 30 - August 4, 2007, Dubna, Moscow Region.}\\
 Bogoliubov Laboratory of Theoretical Physics, JINR, Dubna,
  Moscow Region, 141980, Russia; E-mail: zupnik@theor.jinr.ru}
\vspace{1cm}
\end{center}
\begin{abstract}
We construct the Grassmann-analytic gauge superfields in $D{=}3, N{=}5$ 
harmonic superspace using  the SO(5)/U(1)$\times$U(1) harmonics. These gauge 
$N{=}5$ superfields contain  an infinite number of bosonic and fermionic 
fields arising from  decompositions in harmonics and Grassmann coordinates. 
The bosonic sector of this supermultiplet includes the gauge field $A_m$, the 
additional nongauge vector field $B_m$, the scalar field $S$, two 
SO(5)-vector scalar fields and an infinite number of auxiliary fields with 
SO(5) indices.  The nonabelian Chern-Simons-type action in the $N{=}5$ analytic 
harmonic superspace is constructed. This action is also invariant with 
respect to the sixth supersymmetry realized on the $N{=}5$ gauge superfields. 
The component Lagrangian describes the scale-invariant nontrivial 
interactions of the gauge Chern-Simons field $A_m$ with $B_m, S$
and other basic and auxiliary fields. All auxiliary fields can be excluded
from this Lagrangian.
\end{abstract}
\setcounter{equation}0
\section{Introduction}
 
Supersymmetric extensions of the three-dimensional Chern-Simons (CS) theory
were discussed in refs. \cite{Si}-\cite{Schw}. The $N{=}1$ CS theory of the 
spinor gauge superfield  \cite{Si,Scho} was constructed in the $D{=}3, N{=}1$ 
superspace with real coordinates $x^m, \th^\a$, where $m=0, 1, 2$ is the 3D 
vector index and $\a=1, 2$ is the SL(2,R) spinor index. The $N{=}1$ CS action 
can be interpreted as the superspace integral of the Chern-Simons superform 
$\Tr(dA+\sfrac23A^3)$ in the framework of our theory of superfield integral 
forms \cite{ZP1}-\cite{Z6}.

The abelian $N{=}2$ CS action was first constructed in the $D{=}3, N{=}1$ 
superspace \cite{Si}. The nonabelian $N{=}2$ CS action was considered in the 
$D{=}3, N{=}2$ superspace in terms of the Hermitian superfield $V(x^m,\th^\a,
\bt^\a)$ (prepotential) \cite{ZP1,Iv,NG}, where $\th^\a$ and $\bt^\a$ are the 
complex conjugated $N=2$ spinor coordinates. The corresponding component 
Lagrangian includes the bosonic CS term and the bilinear terms with fermionic 
and scalar fields without derivatives. The unusual dualized form of the $N{=}2$ 
CS Lagrangian contains the second vector field instead of the scalar field
\cite{NG}.

The $D{=}3, N{=}3$ CS theory was first analyzed by the harmonic-superspace 
method \cite{ZK,Z3}. Note that the off-shell $N{=}3$ and $N{=}4$ vector 
supermultiplets are identical \cite{Z5}, however, the superfield CS action is 
invariant with respect to the $N{=}3$ supersymmetry only. Nevetheless, the 
$N{=}3$ CS equations of motion are covariant under the 4th supersymmetry. The 
field-component form of the $N{=}3$ CS Lagrangian was studied in \cite{KL,Kao}.

The physical fields of the $D{=}3, N{=}8$ vector multiplet and the 
corresponding SYM Lagrangian can be found by a dimensional reduction of the  
$D{=}4, N{=}4$ SYM theory. The algebra of supersymmetric transformations 
closes on the SYM$^8_3$ equations of motion only, so it is not clear how to 
use these fields in a hypothetical supersymmetric generalization of the CS 
theory. The off-shell $D{=}3, N{=}6$ SYM theory arises by a dimensional 
reduction of the  SYM$^3_4$ theory in the SU(3)/U(1)$\times$U(1) harmonic 
superspace \cite{GIOS}. The off-shell $N{=}6$ gauge superfields contain the 
physical fields of the SYM$^8_3$ theory and an infinite number of auxiliary 
fields with the SU(3) indices. The integration measure of the corresponding  
$D{=}3, N{=}6$ analytic superspace has dimension 1, and we do not know how to 
construct the CS theory in this superspace. 

In this paper, we consider the simple $D{=}3, N{=}5$ superspace which cannot 
be obtained by a dimensional reduction of the even coordinate from any 4D 
superspace. The corresponding harmonic superspace (HSS) using the 
SO(5)/U(1)$\times$U(1) harmonics is discussed in Section 2. The 
Grassmann-analytic $D{=}3, N{=}5$ superfields depend on 6 spinor coordinates, 
so the analytic-superspace integral measure is scale-invariant. It is not 
difficult to prove that this measure is also invariant with respect to the 
$D{=}3, N{=}5$ superconformal group.

In Section 3, we introduce three basic gauge superfields in the $D{=}3, N{=}5$ 
harmonic analytic superspace. The superfield formalism of this model is 
nominally similar to the HSS formalism of the $D{=}4, N{=}3$ SYM theory 
\cite{GIOS}, although Grassmann dimensions of two superfield models are 
different. One can construct the Chern-Simons-type (CST) superfield action 
from our three $D{=}3, N{=}5$ gauge superfields which looks as a formal analog 
of the $D=4, N=3$ HSS action. Note that one gauge superfield can be 
composed from two mutually conjugated gauge prepotentials. Our superfield 
CST action is invariant with respect to the sixth supersymmetry 
transformation realized on the $N{=}5$ superfields. 

The field-component structure of our $D{=}3, N{=}6$ CST model is analyzed in 
Section 4. In the abelian case, the basic complex gauge superfield includes 
the gauge field $A_m$, the nongauge vector field $B_m$, the scalar field 
$S$, and the fermionic fields $\mu_\a$ and $\psi_\a$. All other fields of the 
infinite off-shell multiplet carry SO(5) indices, for instance, the basic 
real fields $v^a, \nu^a_\a$ and $S^a$. The component abelian Lagrangian
contains Chern-Simons terms for $A_m$ and $B_m$,  the simple interaction
$S\pa_mB^m$, and the bilinear interactions of other fermionic and bosonic 
fields. All abelian auxiliary fields with more than two SO(5) indices vanish
on-shell, so one can  construct the CST Lagrangian on the finite-dimensional
$N=6$ supermultiplet from our superfield action. The basic abelian fields 
$v^a, \mu^\a, \nu^a_\a, S$ and $S^a$ satisfy the free massless equations,
the abelian solution for $A_m$ is pure gauge, but the solution for $B_m$ is
nontrivial. The SO(5)-vector abelian auxiliary fields can be composed on-shell 
from the derivatives of the free fields $v^a, \nu^a_\a$ and $S^a$.

In the nonabelian version of our CST model, the gauge field $A_m$ interacts 
with other fields of the $N{=}6$ gauge supermultiplet, so we cannot obtain
the pure gauge solution for $A_m$ in this case.

\setcounter{equation}0
\section{$D{=}3, N{=}5$ superspace}
Let us consider the $D{=}3, N{=}5$ superspace coordinates $x^m, \th^\a_a,$
where $m=0, 1, 2$ is the SO(2,1) vector index, $a=1, 2,\ldots 5$ is the vector 
index of the automorphism group SO(5), and $\a=1, 2$ is the spinor index of 
the SL(2,R) group. We use the real traceless or symmetric representations of 
the 3D $g_m$ matrices 
\bea
&&(\g_m)^\ab=\ve^{\a\r}(\g_m)^\b_\r=(\g_m)^{\b\a},\q
(\g_m\g_n)^\b_\a=-\eta_{mn}\d^\b_\a+\ve_{mnp}(\g^p)^\b_\a,
\eea
where $\eta_{mn}=\mbox{diag}(1,-1,-1)$ is the 3D Minkowski metric. One can 
consider the bispinor representation of the 3D coordinates and derivatives:
$x^\ab=(\g_m)^\ab x^m,\q \pa_\ab=(\g^m)_\ab \pa_m.$

The $N{=}5$ spinor derivatives are
\bea
&&D_{a\a}=\pa_{a\a}+i\th^\b_a\pa_\ab,\q\pa_{a\a}\th^\b_b=\d_{ab}\d^\b_\a.
\eea

The SO(5)/U(1)$\times$U(1) vector harmonics can be defined as components of the 
real orthogonal matrix
\be
U^K_a=\left(U^{(1,1)}_a, U^{(1,-1)}_a, U^{(0,0)}_a, U^{(-1,1)}_a, U^{(-1,-1)}_a
\right)
\ee
where $a$ is the SO(5) vector index and $K=1, 2,\ldots 5$ corresponds to given 
combinations of the U(1)$\times$U(1) charges. These harmonics satisfy the 
following conditions:
\bea
&&U^K_aU^L_a=g^{KL}=g^{LK},\q g^{KL}U^K_aU^L_b=\d_{ab},\lb{Ubasic}\\
&& g^{15}=g^{24}=g^{33}=1,\q g^{11}=g^{12}=\cdots=g^{45}=g^{55}=0,\nn
\eea
where $g^{LK}$ is the antidiagonal symmetric constant metric in the  space of 
charged indices.

In accordance with a general harmonic approach \cite{GIOS},
we introduce the following harmonic derivatives:
\bea
&&\pa^{KL}=U^K_ag^{LM}\frac{\pa}{\pa U^M_a}-U^L_ag^{KM}\frac{\pa}{\pa U^M_a}=
-\pa^{LK},\\
&&[\pa^{IJ},\pa^{KL}]=g^{JK}\pa^{IL}+g^{IL}\pa^{JK}-g^{IK}\pa^{JL}-g^{JL}
\pa^{IK},
\eea
which form generators of the SO(5) Lie algebra. For instance, the three 
charged harmonic derivatives
\bea
&&\pa^{12}=\pa^{(2,0)}=U^{(1,1)}_a\pa/\pa U^{(-1,1)}_a-U^{(1,-1)}_a\pa/
\pa U^{(-1,-1)}_a,\nn\\
&&\pa^{13}=\pa^{(1,1)}=U^{(1,1)}_a\pa/\pa U^{(0,0)}_a-U^{(0,0)}_a\pa/
\pa U^{(-1,-1)}_a,\nn\\
&&\pa^{23}=\pa^{(1,-1)}=U^{(1,-1)}_a\pa/\pa U^{(0,0)}_a-U^{(0,0)}_a\pa/
\pa U^{(-1,1)}_a
\eea
satisfy the commutation relation
\be
[\pa^{(1,-1)},\pa^{(1,1)}]=\pa^{(2,0)}.
\ee

The Cartan  charges of two U(1) groups are described by the neutral harmonic 
derivatives
\bea
&&\pa^0_1=\pa^{15}+\pa^{24},\q \pa^0_1U^{(p,q)}_a=pU^{(p,q)}_a,\nn\\
&& \pa^0_2=\pa^{15}-\pa^{24},\q \pa^0_2U^{(p,q)}_a=qU^{(p,q)}_a.
\eea

These charges arise in commutators of derivatives with opposite charges, for 
instance,
\bea
&&[\pa^{13},\pa^{35}]=[\pa^{(1,1)},\pa^{(-1,-1)}]=\pa^{15}=\sfrac12(\pa^0_1
+\pa^0_2),\nn\\
&&[\pa^{23},\pa^{34}]=[\pa^{(1,-1)},\pa^{(-1,1)}]=\pa^{24}=\sfrac12(\pa^0_1
-\pa^0_2).
\eea

Let us define the harmonic projections of the $N{=}5$ Grassmann coordinates
\bea
&&\th^K_\a=\th_{a\a}U^K_a=\left(\th^{(1,1)}_\a, \th^{(1,-1)}_\a, \th^{(0,0)}_\a,
\th^{(-1,1)}_\a, \th^{(-1,-1)}_\a\right).
\eea

One can exclude two Grassmann coordinates $\th_\b^{(-1,-1)}$ and 
$\th_\b^{(-1,1)}$ and define the $N=5$ analytic superspace $\zeta=( x^m_\A, 
\th^{(1,1)}_\a, \th^{(1,-1)}_\a, \th^{(0,0)}_\a)$ with 
only three spinor coordinates and the shifted vector coordinate
\bea
&&x^m_\A=x^m+i(\th^{(1,1)}\g^m\th^{(-1,-1)})+i(\th^{(1,-1)}\g^m\th^{(-1,1)}).
\eea

The  general superfields in the analytic coordinates depend also on
additional spinor coordinates $\th^{(-1,1)}_\a$ and $\th^{(-1,-1)}_\a$.

The harmonized partial spinor derivatives are
\bea
&&\pa^{(-1,-1)}_\a=\pa/\pa\th^{(1,1)\a},\q \pa_\a^{(-1,1)}=
\pa/\pa\th^{(1,-1)\a},\q\pa^{(0,0)}_\a=\pa/\pa\th^{(0,0)\a},\lb{partspin}\\
&&\pa^{(1,1)}_\a=\pa/\pa\th^{(-1,-1)\a},\q\pa^{(1,-1)}_\a=\pa/\pa
\th^{(-1,1)\a}.\nn
\eea

An ordinary complex conjugation connects harmonics of the opposite U(1) 
charges
\be
\overline{U_a^{(1,1)}}=U_a^{(-1,-1)},\q\overline{U_a^{(1,-1)}}=U_a^{(-1,1)},
\q\overline{U_a^{(0,0)}}=U_a^{(0,0)}.
\ee
We use mainly the combined conjugation $\sim~$ in the harmonic superspace
\bea
&&\widetilde{U^{(p,q)}_B}=U^{(p,-q)},\q \widetilde{\th^{(p,q)}_\a}=
\th^{(p,-q)}_\a,\nn\\
&&(\th^{(p,q)}_\a\th^{(s,r)}_\b)^\sim=\th^{(s,-r)}_\b\th^{(p,-q)}_\a,\q
\widetilde{f(x_\A)}=\bar{f}(x_\A),
\eea
where $\bar{f}$ is an ordinary complex conjugation.

One can  define the combined conjugation for 
the harmonic derivatives, for instance,
\bea
&&(\pa^{(\pm1,1)}A)^\sim=
\pa^{(\pm1,-1)}\tilde{A},\q(\pa^{(\pm2,0)}A)=-\pa^{(\pm2,0)}\tilde{A}.
\eea

The analytic integral measure contains partial derivatives on the analytic 
spinor coordinates \p{partspin}
\bea
&&d\mu^{(-4,0)}=-\frac{1}{64}d^3x_\A (\pa^{(-1,-1)})^2(\pa^{(-1,1)})^2
(\pa^{(0,0)})^2=d^3x_\A d^6\th^{(-4,0)},\\
&&\int  d^6\th^{(-4,0)}(\th^{(1,1)})^2(\th^{(1,-1)})^2(\th^{(0,0)})^2=1.\nn
\eea
It is pure imaginary
\be
(d\mu^{(-4,0)})^\sim=-d\mu^{(-4,0)},\q(d^6\th^{(-4,0)})^\sim=-d^6\th^{(-4,0)}.
\ee

The harmonic and spinor derivatives can be rewritten in the analytic 
coordinates
\bea
&&\cD^{(1,1)}=\pa^{(1,1)}-i(\th^{(1,1)}\g^m\th^{(0,0)})\pa_m
-\th^{(0,0)\a}\pa^{(1,1)}_\a+\th^{(1,1)\a}\pa^{(0,0)}_\a,\nn\\
&&\cD^{(1,-1)}=\pa^{(1,-1)}-i(\th^{(1,-1)}\g^m\th^{(0,0)})\pa_m-
\th^{(0,0)\a}\pa^{(1,-1)}_\a+\th^{(1,-1)\a}\pa^{(0,0)}_\a,\\
&&\cD^{(2,0)}=[\cD^{(1,-1)},\cD^{(1,1)}]=\pa^{(2,0)}
-2i(\th^{(1,1)}\g^m\th^{(1,-1)})\pa_m-\th^{(1,-1)\a}\pa^{(1,1)}_\a
+\th^{(1,1)\a}\pa^{(1,-1)}_\a,\nn\\
&&D^{(-1,-1)}_\a=\pa^{(-1,-1)}_\a+2i\th^{(-1,-1)\b}\pa_\ab,\q D^{(-1,1)}_\a=
\pa^{(-1,1)}_\a+2i\th^{(-1,1)\b}\pa_\ab,\nn\\
&&
D^{(0,0)}_\a=\pa^{(0,0)}_\a+i\th^{(0,0)\b}\pa_\ab,\q D^{(1,1)}_\a=
\pa^{(1,1)}_\a,\q D^{(1,-1)}_\a=\pa^{(1,-1)}_\a.
\eea

The analytic superfields $\L(\zeta,U)$ depend on harmonics and 
the analytic coordinates and satisfy the conditions
\be
D^{(1,\pm1)}_\a\L=0.
\ee
The harmonic derivatives $\cD^{(1,\pm1)}, \cD^{(2,0)}$ commute with 
$D^{(1,\pm1)}_\a$ and preserve the Grassmann analyticity. 

\subsection{$N{=}5$ superconformal transformations}
The superconformal $D{=}3, N{=}5$ transformations can be defined by analogy 
with the corresponding $D{=}4$ HSS superconformal transformations \cite{GIOS}.
For instance, the  special conformal K-transformations in the $N{=}5$
analytic superspace are
\bea
&&\d_k\, x^\ab_\A=k_{\g\r}x^{\a\g}_\A x^{\b\r}_\A,\q
\d_k\,\th^{(0,0)\a}=x^{\a\b}_\A\th^{(0,0)\g}k_{\b\g},\nn\\
&&\d_k\,\th^{(1,\pm1)\a}=x^{\a\b}_\A\th^{(1,\pm1)\g}k_{\b\g}+\sfrac{i}2\tooS 
\th^{(1,\pm1)\b}k^\a_\b,
\eea
where $k_\ab=(\g^m)_\ab k_m$ are the corresponding parameters.

The K- and SO(5) transformations of harmonics have a similar form 
\bea
&&(\d_k+\d_\o) U^{(0,0)}_a=-\l^{(1,1)}U^{(-1,-1)}_a-\l^{(1,-1)}U^{(-1,1)}_a,
\q(\d_k+\d_\o) U^{(-1,\pm1)}_a=0,\nn\\
&&(\d_k+\d_\o) U^{(1,\pm1)}_a=\l^{(1,\pm1)}U^{(0,0)}_a+\l^{(2,0)}
U^{(-1,\pm1)}_a,\lb{KU}
\eea
where
\bea
&&\l^{(1,\pm1)}=2ik_\ab\th^{(1,\pm1)\a}\th^{(0,0)\b}+U^{(1,\pm1)}_a
U^{(0,0)}_b\o_{ab},\q
\l^{(2,0)}=\cD^{(1,-1)}\l^{(1,1)},
\eea
and $\o_{ab}$ are the SO(5) parameters. 

The S-supersymmetry transformation can be obtained via the Lie bracket
of the special conformal and the Q-supersymmetry transformations
$\d_\eta=[\d_k,\d_\eps]$, where
\bea
&&\d_\eps x^m_\A=-iU_a^{(0,0)}(\eps_a\g^m\th^{(0,0)})
-2iU^{(-1,1)}_a(\eps_a\g^m\th^{(1,-1)})
\nn\\
&&-2iU_a^{(-1,-1)}(\eps_a\g^m\th^{(1,1)}),\qq
\d_\eps \th^{(p,q)\a}=U^{(p,q)}_a\eps^\a_a.
\eea

It is easy to see that the analytic-superspace integral measure
$d^3x_\A d^6\th^{(-4,0)}dU$ is invariant with respect to these
superconformal transformations. 

\setcounter{equation}0
\section{Chern-Simons-type model in $N{=}5$ analytic\\ superspace}

Using a formal analogy with the $D{=}4, N{=}3$ HSS \cite{GIOS} we introduce
the $D{=}3, N{=}5$ analytic matrix gauge prepotentials for our harmonic 
derivatives
\bea
&&V^{(1,1)}(\zeta,U),\q V^{(1,-1)}(\zeta,U),\q V^{(2,0)}(\zeta,U).
\eea
The reality conditions for these prepotentials are
\bea
&&(V^{(1,1)})^\dagger=-V^{(1,-1)},\q (V^{(2,0)})^\dagger=V^{(2,0)},
\eea
where the Hermitian conjugation $\dagger$ includes the $\sim$-conjugation 
of the matrix elements. The infinitesimal gauge transformations of these 
gauge superfields depend on the analytic anti-Hermitian gauge parameter $\L$
\bea
\d_\L V^{(1,\pm1)}=D^{(1,\pm1)}\L+[V^{(1,\pm1)},\L],\q
\d_\L V^{(2,0)}=D^{(2,0)}\L+[V^{(2,0)},\L].
\eea

The covariant harmonic derivatives preserving the G-analyticity have the 
following form:
\bea
&&\na^{(1,\pm1)}=\cD^{(1,1)}+V^{(1,\pm1)},\q 
\na^{(2,0)}=\cD^{(2,0)}+V^{(2,0)},
\eea

The analytic CS-type action can be constructed in terms of three gauge 
prepotentials
\bea
&&S=\frac{i}{g^2}\int dU d^3x_\A d^6\th^{(-4,0)}\Tr\{V^{2,0}\left(\cD^{(1,1)}
V^{(1,-1)}-\cD^{(1,-1)}V^{(1,1)}\right)\nn\\
&&+V^{1,1}\left(\cD^{(1,-1)}V^{(2,0)}-\cD^{(2,0)}V^{(1,-1)}\right)+
V^{1,-1}\left(\cD^{(1,1)}V^{(2,0)}-\cD^{(2,0)}V^{(1,1)}\right)\nn\\
&&-2V^{2,0}[V^{(1,-1)},V^{(1,1)}]+V^{(2,0)}V^{(2,0)}\},\lb{CSact}
\eea
where $g$ is the dimensionless  coupling constant.

The corresponding superfield  gauge equations of motion are
\bea
&&F^{3,1}=\cD^{(1,1)}V^{(2,0)}-\cD^{(2,0)}V^{(1,1)}+[V^{(1,1)},V^{(2,0)}]=0,
\lb{CSequ}\\
&&V^{(2,0)}=\cD^{(1,-1)}V^{(1,1)}-\cD^{(1,1)}V^{(1,-1)}+[V^{(1.-1)},V^{(1,1)}]
\equiv\hat{V}^{(2,0)}.
\eea

The last superfield can be composed algebraically in terms of two other 
prepotentials. Using the substitution  $V^{(2,0)}\rightarrow \hat{V}^{(2,0)}$ 
in \p{CSact} we can obtain the  alternative form of the  action with only two 
independent prepotentials $V^{1,1}$ and $V^{1,-1}$
\bea
&&S_2=\frac{i}{g^2}\int dU d\mu^{(-4,0)}\Tr\{V^{1,-1}\cD^{(2,0)}V^{(1,1)}
-V^{1,1}\cD^{(2,0)}V^{(1,-1)}
\nn\\
&&-\left(\cD^{(1,-1)}V^{(1,1)}-\cD^{(1,1)}V^{(1,-1)}+[V^{(1,-1)},V^{(1,1)}]
\right)^2\}.\lb{CSact2}
\eea

It is evident that the superfield action \p{CSact} is invariant with respect 
to the sixth sypersymmetry transformation defined on our gauge potentials
\bea
\d_6\left(V^{(1,\pm1)},V^{(2,0)}\right)=\eps^\a_6D^{(0,0)}_\a\left(
V^{(1,\pm1)},V^{(2,0)}\right),
\eea
where $\eps^\a_6$ are the corresponding spinor  parameters. Thus, our
superfield CST gauge model possesses the $D{=}3, N{=}6$ supersymmetry. 

The CST actions \p{CSact} and \p{CSact2} are formally similar to the HSS 
actions of the  SYM$_4^3$ theory\cite{GIOS} (or to the action of the
dimensionally reduced SYM$^6_3$ theory), although the Grassmann dimensions of 
analytic superspaces in these two types of models are different. In the next 
section, we analyze the field-component structure of our superfield 
model. 

\setcounter{equation}0
\section{Harmonic component fields in the $N{=}6$  Chern-Simons-type model}

The abelian form of the action $S_2$ \p{CSact2} contains $\sim$-conjugated
abelian prepotentials
\bea
&&S_2^0=i\int dU d\mu^{(-4,0)}\{V^{1,-1}\cD^{(2,0)}V^{(1,1)}
-V^{1,1}\cD^{(2,0)}V^{(1,-1)}\nn\\
&&-\left(\cD^{(1,-1)}V^{(1,1)}-\cD^{(1,1)}V^{(1,-1)}
\right)^2\}.\lb{CSact20}
\eea

The Grassmann and harmonic decompositions of the imaginary gauge superfield
parameter have the following form:
\bea
&&\L(\zeta,U)=i[a+U^{(0,0)}_ba^b]
+\th^{(1,1)\a}U^{(-1,-1)}_b\r_\a^b
+\th^{(1,-1)\a}U^{(-1,1)}_b\bar\r_\a^b
+\th^{(0,0)\a}[\b_\a+U^{(0,0)}_b\b^b_\a]\nn\\
&&+(\th^{(0,0)})^2[d+U^{(0,0)}_ad_a]
+(\th^{(1,1)}\th^{(0,0)})U^{(-1,-1)}_bf^{b}+
(\th^{(1,-1)}\th^{(0,0)})U^{(-1,1)}_b\bar{f}^{b}
\nn\\
&&
+(\th^{(1,1)\a}\th^{(0,0)\b})U^{(-1,-1)}_bg^{b}_\ab
+(\th^{(1,-1)\a}\th^{(0,0)\b})U^{(-1,1)}_b\bar{g}^{b}_\ab\nn\\
&&+(\th^{(0,0)})^2\th^{(1,1)\a}U^{(-1,-1)}_b\pi_\a^b
-(\th^{(0,0)})^2\th^{(1,-1)\a}U^{(-1,1)}_b\bar\pi_\a^b
+O(U^2),
\eea 
where all coefficients depend on $x_\A^m$. Bilinear in harmonics terms 
and the corresponding $\th$ terms are omitted, and 
the  condition $\L^\sim=-\L$ is used in this formula.

The pure gauge degrees of freedom in the prepotential $V^{(1,1)}$ can be 
eliminated by the transformation $\d V^{(1,1)}=\cD^{(1,1)}\L$. In the 
WZ-gauge, the superfield parameter is reduced to $\L_{WZ}=ia(x_\A)$, but 
the corresponding gauge superfield still has an infinite number of 
component fields
\bea
&&V^{(1,1)}_{WZ}=U^{(1,1)}_av^a+i\th^{(1,1)\a}\mu_\a
+i\th^{(0,0)\a}U^{(1,1)}_a\nu^a_{\a}-(\th^{(1,1)}\g^m\th^{(0,0)})(A_m+iB_m)
\nn\\
&&+(\th^{(1,1)}\g^m\th^{(1,-1)})U^{(-1,1)}_aC^a_m+i(\th^{(1,1)}\th^{(0,0)})
(S+U^{(0,0)}_aS^a)+\tooS U^{(1,1)}_a b^a\nn\\
&&+(\th^{(1,1)})^2U^{(-1,-1)}_a h^a+\th^{(1,1)\a}\th^{(1,-1)\b}\th^{(0,0)\g}
U^{(-1,1)}_a\J^a_{(\a\b\g)}+(\th^{(1,-1)}\th^{(0,0)})\th^{(1,1)\a}
U^{(-1,1)}_a\xi^a_\a\nn\\
&&+(\th^{(1,1)}\th^{(0,0)})\th^{(1,-1)\a}U^{(-1,1)}_a\l^a_\a
+(\th^{(0,0)})^2\th^{(1,1)\a}[\j_\a+U^{(0,0)}_a\j^a_\a]\nn\\
&&+(\th^{(1,1)})^2(\th^{(0,0)})^2U^{(-1,-1)}_aF^a+(\th^{(1,1)}\th^{(1,-1)})
(\th^{(0,0)})^2U^{(-1,1)}_aG^a\nn\\
&&+(\th^{(1,1)}\g^m\th^{(1,-1)})(\th^{(0,0)})^2
U^{(-1,1)}_aG^a_m+O(U^2),\lb{VWZ}
\eea
where again bilinear in harmonics terms and the corresponding Grassmann terms
are omitted for brevity. The basic fields of this supermultiplet
\be
v^a, \mu_\a, \nu_\a^a, A_m, B_m, S, S^a\lb{basic}
\ee
are real and other fields are complex. Note that $A_m$ is the gauge field 
($\d A_m=\pa_ma$), but we have no gauge parameter for the second vector 
field $B_m$. Dimensions of these off-shell fields are
\bea
&&[v^a]=0,\q[\mu_\a]=[\nu^a_\a]=\sfrac12,\q [A_m]=[B_m]=[S]=[S^a]=[b^a]=
[C^a_m]=1,\nn\\
&&[\psi_\a]=[\psi^a_\a]=[\xi^a_\a]=[\l^a_\a]=[\J^a_{\a\b\g}]=\sfrac32,\q
[F^a]=[G^a]=[G^a_m]=2.
\eea
In the representation \p{VWZ}, we omit an infinite number of auxiliary
bosonic and fermionic fields $F^{a_1\ldots a_k}\q (k\geq 2)$ with multiple 
vector SO(5) indices.

The component Lagrangian for the SO(5) invariant fields has the following 
form:
\bea
&&L_0=3\ve^{mnr}A_m\pa_nA_r+\ve^{mnr}B_m\pa_nB_r-4S\pa^mB_m-i\j^\a\j_\a
-i\bj^a\bj_\a+i\j^\a\bj_\a\nn\\
&&+\sfrac{3i}2(\j^\a+\bj^\a)\pa_\ab\mu^\b-\sfrac{i}4\pa^\a_\b\mu^\b
\pa_{\a\g}\mu^\g.\lb{L0}
\eea
The corresponding equations of motion have the pure gauge solution for $A_m$ 
and the nontrivial solution for the  vector field $B_m$. The scalar field $S$ 
and the noncanonical fermionic field $\mu^\a$ satisfy
the free equations $\pa^m\pa_mS=0,\q\pa^m\pa_m\mu^\a=0$. 
The on-shell construction  for the auxiliary fermionic field is 
$\j_\a=\sfrac32\pa_\ab\mu^\b$.

It is not difficult to construct the Lagrangian for the SO(5) vector fields 
from \p{VWZ} using the superfield action \p{CSact20}. The basic SO(5) vector 
fields satisfy the free equations
$\pa^m\pa_mv^a=\pa^m\pa_mS^a=\pa^m\pa_m\nu^\a_\a=0,$ 
and the on-shell complex auxiliary fields can be composed from these basic real
fields
\bea
&& b^a=-\sfrac{3i}2S^a,\q h^a=2iS^a,\q C^a_m=3i\pa_mv^a,\q F^a=-\sfrac12
\pa^m\pa_mv^a,\q G^a=-\sfrac13\pa^m\pa_mv^a,\nn\\
&& G^a_m=\pa_mS^a,\q\xi^a_\a=\pa_\ab\nu^{a\b},\q\l^a_\a=-\sfrac53\pa_\ab
\nu^{a\b},\q\j^a_\a=\sfrac43\pa_\ab\nu^{a\b},\q\J^a_{\a\b\g}=
-\pa_{(\ab}\nu^a_{\g)}.\nn
\eea
All abelian auxiliary fields with more than two SO(5) indices vanish on-shell.

Thus, the solutions for the superfield N=5 gauge prepotentials contain 
the nontrivial vector,  scalar and fermion fields in distinction with the pure 
gauge superfield solutions of the $N{=}1, 2$ and 3 supersymmetric Chern-Simons 
theories. 

We plan to analyze the nonabelian interactions of the Chern-Simons gauge field 
$A_m$ with the fields $v^a, \mu_\a, \nu^a_\a, B_m, S$ and $S^a$ (all fields in 
the adjoint representation of the gauge group) using the algebraic auxiliary 
field equations in our $N{=}6$ Chern-Simons-type model. The superfield 
representation may be useful for quantum calculations. 

I am grateful to E.A. Ivanov for  interesting discussions.
This work was partially supported by  DFG grant 436 RUS 113/669-3 , by
RFBR grants 06-02-16684 and 06-02-04012, by  INTAS grant 05-10000008-7928
and by  grant of the Heisenberg-Landau programme.

\end{document}